\documentclass[a4paper]{spie}  

\usepackage{amsmath,amsfonts,amssymb}
\usepackage{graphicx}
\usepackage[colorlinks=true, allcolors=blue]{hyperref}

\title{The evaluation of the CUSP scientific performance
by a GEANT4 Monte Carlo simulation}

\author[e]{Giovanni~De~Cesare}
\author[a]{Sergio~Fabiani}
\author[e]{Riccardo~Campana}
\author[a,i]{Giovanni~Lombardi}
\author[a]{Ettore~Del~Monte}
\author[a]{Enrico~Costa}
\author[b]{Ilaria~Baffo}
\author[c]{Sergio~Bonomo}
\author[g]{Daniele~Brienza}
\author[h]{Mauro~Centrone}
\author[d]{Gessica~Contini}
\author[c]{Giovanni~Cucinella}
\author[f]{Andrea~Curatolo}
\author[a]{Nicolas~De~Angelis}
\author[d]{Andrea~Del~Re}
\author[a]{Sergio~Di~Cosimo}
\author[c]{Simone~Di~Filippo}
\author[a]{Alessandro~Di~Marco}
\author[a]{Giuseppe~Di~Persio}
\author[g]{Immacolata~Donnarumma}
\author[b]{Pierluigi~Fanelli}
\author[d]{Paolo~Leonetti}
\author[f]{Alfredo~Locarini}
\author[a]{Pasqualino~Loffredo}
\author[a]{Gabriele~Minervini}
\author[f]{Dario~Modenini}
\author[a]{Fabio~Muleri}
\author[g]{Silvia~Natalucci}
\author[c]{Andrea~Negri}
\author[c]{Massimo~Perelli}
\author[a]{Monia~Rossi}
\author[a]{Alda~Rubini}
\author[a]{Emanuele~Scalise}
\author[a]{Paolo~Soffitta}
\author[g]{Andrea~Terracciano}
\author[f]{Paolo~Tortora}
\author[g]{Emauele~Zaccagnino}
\author[d]{Alessandro~Zambardi}

\affil[a]{INAF-IAPS, via del Fosso del Cavaliere 100, 00133 Rome, Italy}
\affil[b]{DEIM, Universit\'a degli studi della Tuscia, Largo dell'Universit\'a, 01100 Viterbo, Italy}
\affil[c]{IMT s.r.l., via Carlo Bartolomeo Piazza 30, 00161 Rome, Italy}
\affil[d]{SCAI Connect s.r.l.,  Via Francesco Gentile 135, 00173 Rome, Italy}
\affil[e]{INAF-OAS Bologna, via Piero Gobetti 93/3, 40129 Bologna, Italy}
\affil[f]{Alma Mater Studiorum Universit\'a di Bologna - Department of Industrial Engineering and Interdepartmental Center for Industrial Aerospace Research, Via Fontanelle 40, 47121 Forl\'i, Italy}
\affil[g]{ASI, Via del Politecnico snc 00133 - Roma, Italy}
\affil[h]{INAF-OAR, Via Frascati 33, 00040, Monte Porzio Catone, Italy}
\affil[i]{ Dipartimento di Ingegneria dell’Impresa ''Mario Lucenti", Università degli Studi di Roma ``Tor Vergata”, Via Cracovia 50, 00133 Roma, Italy}

\authorinfo{Further author information: (Send correspondence to Giovanni De Cesare)\\Giovanni De Cesare: E-mail: giovanni.decesare@inaf.it, Telephone: +39 051 6398776}

\begin{document} 
\maketitle

\begin{abstract}
The CUbesat Solar Polarimeter (CUSP) project is a CubeSat mission orbiting the Earth aimed to measure the linear polarization of solar flares in the hard X-ray band by means of a Compton scattering polarimeter. CUSP will allow to study the magnetic reconnection and particle acceleration in the flaring magnetic structures of our star. CUSP is a project in the framework of the Alcor Program of the Italian Space Agency aimed to develop new CubeSat missions. It is approved for a Phase B study. In this work, we report on the accurate simulation of the detector's response to evaluate the scientific performance. A GEANT4 Monte Carlo simulation is used to assess the physical interactions of the source photons with the detector and the passive materials. Using this approach, we implemented a detailed CUSP Mass Model. In this work, we report on the evaluation of the detector's effective area as a function of the beam energy.
\end{abstract}


\section{INTRODUCTION}
\label{sec:intro}  
The CUbesat Solar Polarimeter (CUSP) project [\citenum{Fabiani2022, fabiani2024cusp}] aims to measure the linear polarisation of solar flares in the hard X-ray band by means of a Compton
scattering polarimeter on board of a constellation of two CubeSats. CUSP is going
to start the Phase B in 2024 funded by the Alcor program of the Italian Space
Agency. INAF-IAPS is Prime and scientific Principal investigator of the project. In
this work we introduce the GEANT4 [\citenum{Agostinelli2003}] C++ simulation code of the polarimeter's
response under development. The first parameter under assessment is a precise
estimation of the detector's effective area. The same code will be an essential tool
for a deeper estimation of the CUSP scientific performance, based on the
estimation of the modulation factor, the spurious modulation and the instrument
background. Moreover, it will support the development and test of the data
analysis software.

\section{THE SIMULATION CODE}
\label{sec:simulation_code}
\begin{figure} [ht]
   \begin{center}
   \begin{tabular}{c} 
   \includegraphics[height=5cm]{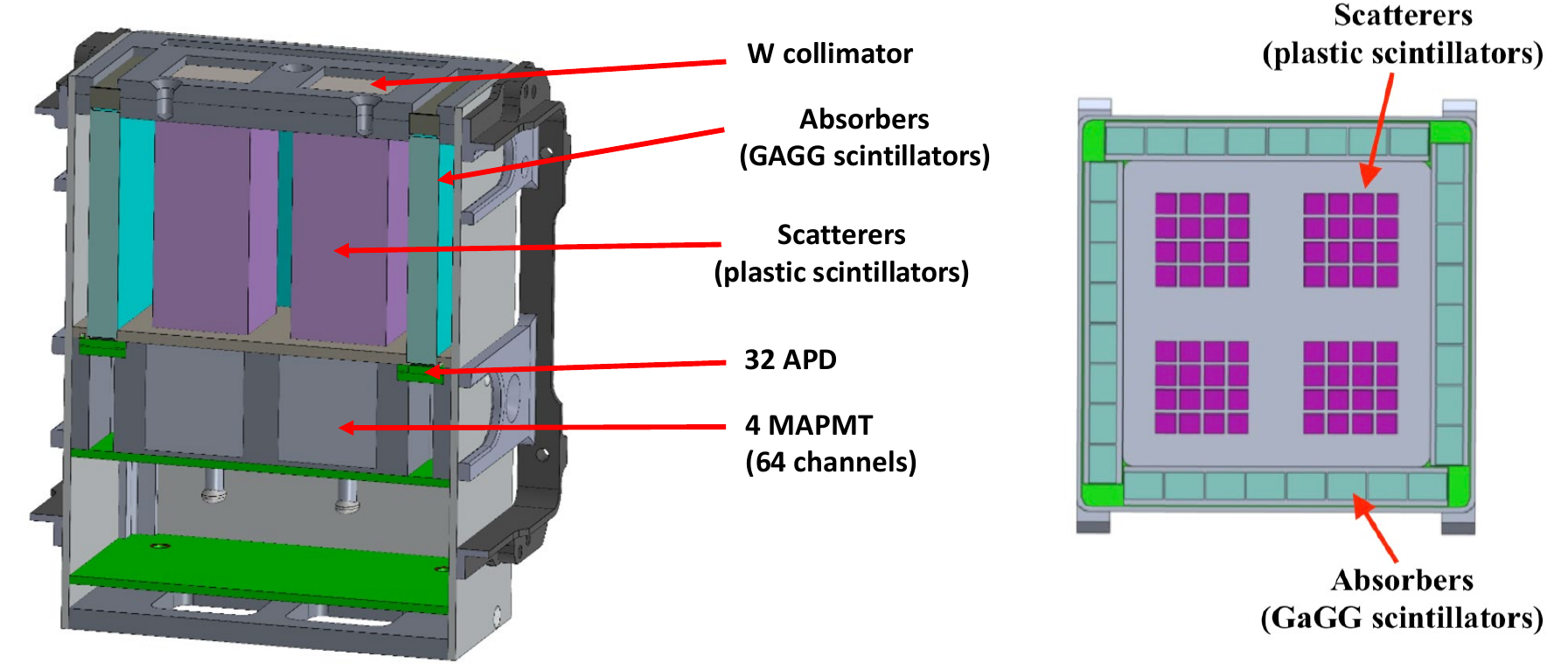}
   \end{tabular}
   \end{center}
   \caption[cad] 
   { \label{fig:cad} 
   Scheme of CUSP X-ray polarimeter. Left panel: Cut side view. Right panel: top view.
}
   \end{figure}

A GEANT4 Monte Carlo framework is used to assess the physical interactions of the hard X-ray photons with the detector and the passive materials. Using this approach,
we implemented a detailed CUSP Mass Model. It is derived from a simplified CAD
model converted into a GDML model, then directly loaded in the GEANT4 simulation
framework. framework. The CAD, and thus the GDML discussed in this work, is the one assessed
in \citenum{lombardi2024cusp} . CUSP (Fig. \ref{fig:cad}) is a dual-phase Compton diffusion polarimeter. The incident X-rays are scattered by plastic scintillator and absorbed by crystals of inorganic
scintillator (GAGG, gadolinium aluminum gallium garnet).

   \begin{figure} [ht]
   \begin{center}
   \begin{tabular}{c} 
   \includegraphics[height=6cm]{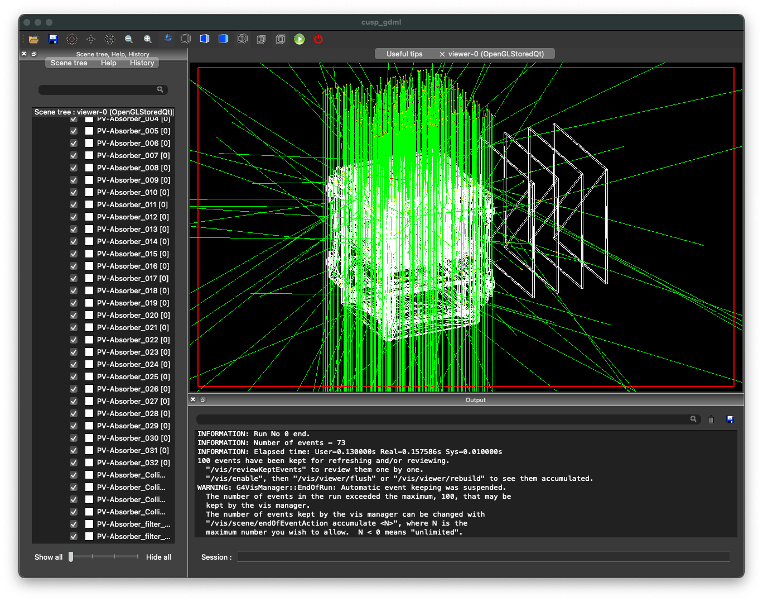}
   \end{tabular}
   \end{center}
   \caption[cad] 
   { \label{fig:geant4} 
   The GEANT4 GUI of the simulation of a parallel beam impinging on the CUSP polarimeter. The 60 keV photons of the simulated source are visualized as green lines.
}
   \end{figure}

The Monte Carlo simulation can be launched in interactive or in batch mode. The
interactive mode (fig. \ref{fig:geant4}) is used to visualize the mass model in detail and trace the
physical interaction of photons with the materials. This enables us to debug and test
the code. In batch mode, it is possible to generate a large simulation and obtain as an
output a file of events, ready to be analyzed. This file associates each photon from
the beam with the list of the sensitive volumes impacted by an energy deposit.

\section{ESTIMATION OF THE EFFECTIVE AREA}
\label{sec:eff_area}

   \begin{figure} [ht]
   \begin{center}
   \begin{tabular}{c} 
   \includegraphics[height=8cm]{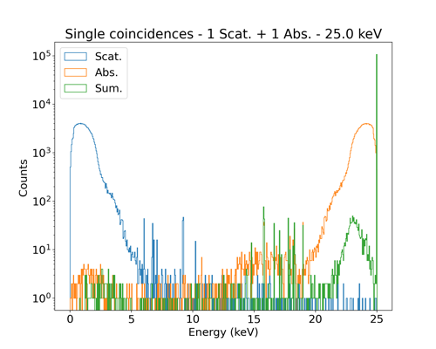} 
   \end{tabular}
   \end{center}
   \caption[cad] 
   { \label{fig:fig3} 
   Spectrum of the energy deposits for the one scatterer and one absorber coincident events.
}
   \end{figure}
   
\begin{figure} [ht]
   \begin{center}
   \begin{tabular}{c} 
      \includegraphics[height=7cm]{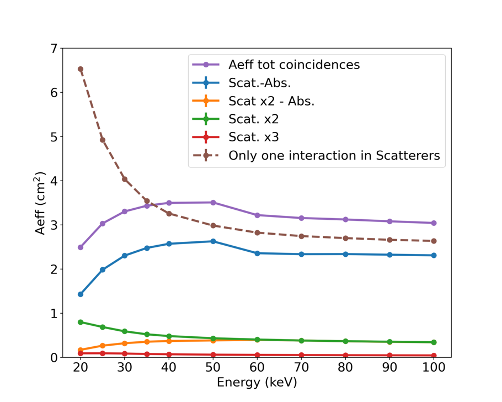}
   \includegraphics[height=7cm]{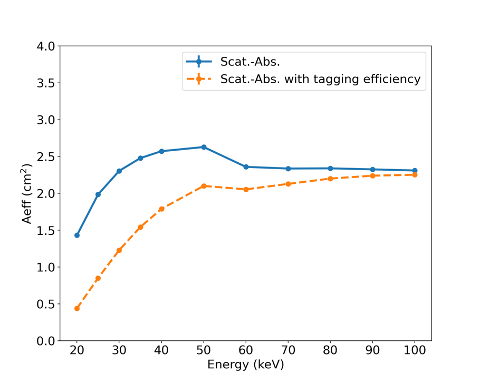}
   \end{tabular}
   \end{center}
   \caption[cad] 
   { \label{fig:fig4} 
   \textit{Left panel}: CUSP effective area versus energy for different families of events. Scat.-Abs: one energy deposits in the scatterer and one energy deposits in the absorber. Scat.~x2-Abs: two energy deposits in the scatterer, one energy deposit in the absorber.  Scat.~x2: two energy deposits in the scatterer. Scat.~x3: three energy deposits in the scatterer.  \textit{Right panel}: CUSP effective area as a function of energy, corrected by tag efficiency (orange line). 
}
   \end{figure}

When an X-ray photon is detected by the polarimeter, the Compton energy
deposits can occur in one or more scintillating bars. Polarimetric measurements
require the coincident measurement in one or two scatterer bars plus a an
absorber. Also a measurement of a coincidence in two scatterers is suitable to
measure an azimuthal scattering angle. Figure \ref{fig:fig3} shows the simulated spectrum
of energy deposits in the scatterers, in the absorbers, and the sum of the two
energy deposits for a mono-energetic beam at 25 keV producing 1-scatterer/1-
absorber coincidences. 
Left panel of Figure \ref{fig:fig4} shows the estimate of the effective area for different families of coincidences. Coincidences between a scatterer and an absorber (blue curve) are the larger contribution to the effective area. Right panel compares the ideal case (blue line) and by applying the so called tagging efficiency (orange line) that accounts for the probability to detect a faint light signal in the scatterer once the signal in the absorber was detected. The estimation of the tagging efficiency is from [\citenum{Fabiani2012c}] for a similar (but not the same) set-up of plastic bar coupled to a photo-multiplier tube.

\section{CONCLUSIONS}
\label{sec:conclusions}
With the aim of obtaining an accurate evaluation of the
scientific performance of CUSP, we developed a Monte
Carlo code based on GEANT4, starting from a detailed
mass model of the instrument. The same code will be used
to evaluate the polarimetric response of the instrument
and will also be a useful tool for calibrations and the
development of data analysis software. As a first result, the
estimation of effective areas for different event definitions
has been shown in this work.

\bibliography{main} 
\bibliographystyle{spiebib} 

\acknowledgments 
Activity funded by ASI phase A contract 2022-4-R.0.

\end{document}